\documentclass[prd,aps,showpacs,tightenlines,nofootinbib,epsfig]{revtex4}
\usepackage{bm,exscale,latexsym, amsmath, amssymb, mathrsfs, dsfont,tensor}
\usepackage{slashed} 
\usepackage{graphicx}

\newcommand{\bear}{\begin{array}}  \newcommand{\eear}{\end{array}}
\newcommand{\bea}{\begin{eqnarray}}  \newcommand{\eea}{\end{eqnarray}}
\newcommand{\beq}{\begin{equation}}  \newcommand{\eeq}{\end{equation}}
\newcommand{\bef}{\begin{figure}}  \newcommand{\eef}{\end{figure}}
\newcommand{\bec}{\begin{center}}  \newcommand{\eec}{\end{center}}
\newcommand{\bed}{\begin{description}}  \newcommand{\eed}{\end{description}}

\begin{document}

\title{Gravity triggered neutrino condensates}
%
%
%
\author{Gabriela Barenboim}
\affiliation{Departament de F\'{\i}sica Te\`orica and IFIC, Universitat de 
Val\`encia-CSIC, E-46100, Burjassot, Spain.}
\date{\today}
\begin{abstract}
In this work we use the Schwinger-Dyson equations to 
study the possibility that an enhanced gravitational
attraction triggers the formation of a right handed neutrino
condensate, inducing dynamical symmetry breaking and generating a Majorana
mass for the right handed neutrino at a scale appropriate for the see-saw 
mechanism. The composite field formed by the condensate phase could drive an early epoch of inflation.

We find that to the lowest order, the theory does not allow dynamical
symmetry breaking.
Nevertheless, thanks to the large number of matter fields
in the model,  the 
 suppression by additional powers in $G$ of  higher order terms can be
compensated, boosting them up to their lowest order counterparts. This way
 chiral symmetry can be broken dynamically and the infrared mass
generated  turns out to be in the expected range for a successful 
see-saw scenario.
\end{abstract}

\maketitle


\section{Introduction}

\label{sec:introduction}
The most commonly accepted theory about the origin and the evolution of our universe 
is  doubtlessly the hot big bang model.
Such a model is based on two crucial observations: the
discovery of the expansion of the universe as depicted
by the Hubble's law, and the existence of an astonishing
isotropic and a perfectly thermal Cosmic Microwave 
Background (CMB) radiation. Since the energy
density of radiation dilutes away faster with the expansion than
that of matter, these two observations immediately point
to the fact that the universe has evolved from a hot and
dense early phase, when radiation, and not  matter,
was dominant. The recent evolution to a matter
dominated epoch takes place when the radiation density
falls sufficiently low that the photons cease to interact
with matter. The CMB is then nothing but the remnant of a lost epoch,
in which radiation was the dominant component of the energy budget, the relic radiation
which is reaching us today from this epoch of decoupling.
While reasonably isotropic, the CMB possesses small
anisotropies (of about one part in $10^5$), and the observed
pattern of the fluctuations in the CMB provides a straight
snapshot of the universe at this epoch. The hot big bang
model has been rather successful in predicting, say, the
primordial abundances of the light elements in terms of
a single parameter, namely the baryon-to-photon ratio, and
the amount required to fit these observations beautifully agrees with the
value that has been arrived at independently from the
structure of the anisotropies in the CMB. Despite the
success of the hot big bang model in explaining the results
from different observations, the model has a serious
drawback. Under this framework, the CMB photons arriving
at us today from sufficiently widely separated directions in the sky 
could not have interacted at the time of decoupling.
Nevertheless, one finds that the temperature of
the CMB photons reaching us from any two diametrically
opposite directions scarcely vary \footnote{For an account of the 
impressive successes and the scarce weaknesses of the hot big bang model
see, for example \cite{hbb}}.

Inflation \cite{inflation}, which refers to a period of accelerated expansion
during the early stages of the radiation dominated
epoch, provides a satisfactory resolution to abovementioned
deficiencies of the hot big bang model. Actually, in addition to offering a
graceful explanation for
the extent of homogeneity and isotropy of the background
universe, inflation also provides an attractive
causal mechanism to generate the inhomogeneities superimposed
upon it. The inflationary phase blows up
the tiny quantum fluctuations present at the beginning
of the epoch and transforms them into classical perturbations
which leave their watermarks as anisotropies in the
CMB.  Subsequently these anisotropies  act as seeds for the
formation of the large scale structures that we observe
at the present time as galaxies and clusters of galaxies.
With the anisotropies in the CMB being measured
to greater and greater precision, we have an unprecedented
fertile ground to test the predictions of inflation. The simplest models of inflation driven by
a single, slowly rolling scalar field, generically predict a
nearly scale invariant spectrum of primordial perturbations,
which seems to be in excellent agreement with the
recent observations of the CMB.

Although the inflationary paradigm is widely accepted, most attempts to incorporate
inflation into specific models of particle physics struggle with  two weaknesses. 
(i) Parameters
such as coupling constants must often be "fine-tuned"  to extremely small values in
order to avoid massive overproduction of density fluctuations. This can be 
traced back in a general
sense to emanate from an exponential dependence of the vacuum energy density $V^4$ on the symmetry
breaking scale $V$. (ii) Symmetry breaking scales significantly below the Planck scale,
$ V < M_{Pl} \simeq  10^{19} \mbox{GeV}$ 
 are at odds  with observational constraints. The latter deficiency is
particularly bothering, since physics at the Planck scale is at present poorly understood, and
there is not a powerful  basis to assume that standard notions of spontaneous symmetry
breaking are valid at such high energies.

Models giving rise to inflation are typically formulated in terms of an
elementary scalar field, the inflaton. In most models the inflaton field,
$\phi $,
is conceived as a scalar field (associated in the most ambitious models
with a grand unified Higgs field) whose effective potential $V_{eff} (\phi)$
is chosen such that $\phi $ is initially trapped in a local minimum at $\phi =0 $
(where it is specially flat)
and has to roll down slowly  to its true minimum
at $\phi = \phi_0 $. For a suitable choice of the model parameters 
 the universe would spend
a significant period of time in an exponentially expanding de
Sitter phase, with its energy density dominated by the large value
$V_{eff} (0) - V_{eff} (\phi_0)$.
A crucial fine tuning is required in these models. Let alone the fact that
these models reckon upon the existence of a fundamental scalar field.

As fundamental scalar fields are yet to be observed, in this work we
explore the possibility of giving the inflaton a structure, as a dynamically
generated condensate of right handed neutrinos triggered by gravitation.

Why neutrinos? Neutrinos have provided our first (and so far the only)
glimpse beyond the Standard Model as neutrino oscillation experiments
have shown unambiguous evidence that the three active neutrinos have mass
and mix. Neutrino masses, as deduced from oscillation experiments point towards
the existence of right handed neutrino states and a new energy scale.
The new energy scale, is associated with the fact that the triviality of the
quantum numbers of the new states allows them to have Majorana masses of
order $M$, as well as to couple to the $SU(2)_L$ doublets and Higgs bosons.
Although all values for $M$ are technically natural, large values
of $M$ are preferred for several reasons, including the fact that one may relate
$M$ to other well justified high energy scales, as $M_{Pl}$ and $M_{GUT}$.
Within  the see-saw mechanism \cite{seesaw}, 
a theoretically appealing scenario to generate the observed
light masses in a natural way, the coupling to the Higgs doublet becomes a Dirac mass
of order $\mu$ after electroweak symmetry breaking, while the Majorana masses are
at some high scale, $M \sim M_{GUT}$. In this case, the resulting propagating neutrino
degrees of freedom separate into two quasi-decoupled groups: mostly active
states with masses $ m \sim \mu^2/M$ and mostly sterile states with 
very large masses.
In this case the right handed neutrinos will have  super-heavy masses, as compared
to their standard model counterparts. Being so heavy turns them into ideal candidates
to provide structure (dynamical origin)  to  another super-heavy and weakly 
coupled particle, the inflaton. 

Trying to form a right handed neutrino condensate, faces an obvious problem.
Neutrinos are weakly inteacting particles. Even more, right
handed neutrinos are solo players in the standard model, leaving gravity as
their only possibility. Gravity, a force known to be miserably weak
among elementary particles. So we want to explore the prospects of gravitational
interactions between right handed neutrinos at ultra-high energies, i.e. in the very
early universe, being strong enough to trigger
the formation of a low-energy condesate, breaking lepton number and giving masses
to right handed neutrinos. 

As before, we can ask again, why neutrinos ? wouldn't this work (if it indeed
works) for any other fermion as well ? If neutrinos do condensate, why doesn't it
happen to say, electrons or quarks ?
The answer to  this puzzling behaviour rests on the fact that
neutrinos are the only known fundamental neutral fermion. And this fact singles
them out from the rest. For all charged fermions, when gauge interactions are attractive,
a new annihilation channel opens up, closing the door to condensation. On the other hand,
when annihilation is not possible, gauge interactions are repulsive.

\section{Setting the scene}

\label{sec:right}

We will investigate whether chiral symmetry is broken dynamically due to the 
formation of a right handed neutrino condesate triggered by gravity.
The formation of such a condensate, if it extists,
sets off dynamical symmetry breaking of lepton number and produces a Majorana mass for the
right-handed neutrino. The phenomenology of right handed neutrino condensates 
in the early universe was studied in
 \cite{Barenboim:2008ds} by introducing an effective four fermion self-coupling
for the right handed neutrinos. 
It was found there that the condensate dynamics also produces ``natural inflation" \cite{natural}. 
This way both, the inflationary de Sitter scale and the right handed
neutrino mass, come out naturally  of order of the infrared mass of the
condensate. The quantum fluctuations
of the condensate, which mimics a scalar field, give rise to primordial density fluctuations 
of the needed size and a spectral index in agreement with observations. The tensor perturbations
however end up being exponentially suppressed, while the predictions for the running of the spectral index 
come out to be negligible small, experimentally indistinguishable from zero, 
making the model especially easy to test in the next generation
of experiments.
As compared to the usual approach to
inflaton model building a dynamical framework is both economical and predictive.
In this work we will try to relate the dynamics underlying the effective four fermion
interaction to quantum gravity.

The Schwinger-Dyson equations provide a convenient way to study dynamical
symmetry breaking and dynamical mass generation in quantum field theory. However,
they constitute an infinite set of integral equations and some truncation scheme
is necessary in order to explore its possible solutions.

Since quantum gravity is known to be non renormalizable, it will
be necessary to introduce a momentum cut-off $\Lambda$ into the theory. Intuitively one
assumes that $\Lambda $ is of order $M_{Pl}$, Planck's mass. 
We will see that when the number of matter fields is large a lower, reduced Planck
scale, will play an important role.

We can now explore dynamical symmetry breaking using 
the Schwinger-Dyson equation for the fermion propagator.
Because of Lorentz and parity invariance the full propagator for the neutrino 
field $S_F$ can be written  in the massless limit as
\bea
S_F^\prime(p)= \frac{i}{\slashed p - \Sigma(p)}= \frac{i}{\alpha(p^2) \slashed p - \beta(p^2)}
\label{eq:fermionprop}
\eea
then the self energy part $\Sigma(p)$ satisfies 
the following integral equation
\bea
\Sigma (p) = \int \frac{d^4 k}{(2\pi)^4} \;\; \Gamma_{\alpha \beta}(k-p,p) \; S_F(k)
\; G^{\prime \alpha \beta \mu \nu} (p-k) \; \Gamma^\prime _{\mu \nu} (p-k,k)
\eea
\begin{figure}[htb]
\begin{minipage}[c][0.1\textheight][t]{0.45\textwidth}
\includegraphics[scale=.95]{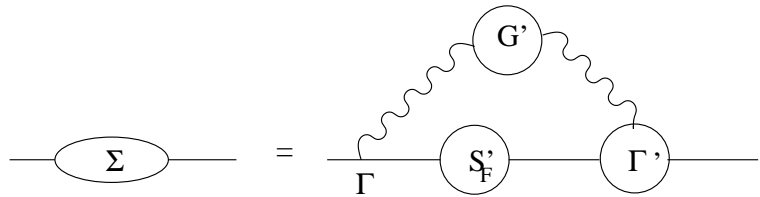}
\end{minipage}
\caption{The Schwinger-dyson equation for the fermion full propagator. Prime denotes full quantities}
\label{self}
\end{figure}
where  $\Gamma^{(\prime)} _{\mu \nu} (p,k)$ is the tree (full) fermion-fermion-graviton
vertex function and $G^{\prime \alpha \beta \mu \nu}$ stands for the full graviton 
propagator. We will perform our calculation in the standard harmonic gauge and 
will consider an approximation where
we replace $G^\prime$ and $\Gamma^\prime$ by their
tree level values $G$ and $\Gamma$ respectively.
Then, for the lowest order,  we obtain the coupled integral  
equations
\bea
\alpha(p^2)= 1- i \kappa^2 \int \frac{d^4 k}{(2\pi)^4} \;\;  \frac{\alpha(k^2)}{\alpha(k^2) 
k^2 - \beta^2 k^2}\frac{1}{16 (p-k)^2 p^2} \left( 4 k^2 p^2 + 3 (k^2 + p^2) kp + 2 (kp)^2 
\right) 
\eea
where $\kappa^2 = 32 \pi G $ and
\bea
\beta(p^2)=  i \kappa^2 \int \frac{d^4 k}{(2\pi)^4} \;\;  \frac{\beta(k^2)}{\alpha(k^2) 
k^2 - \beta^2 k^2} L(p,k)
\eea
Within the approximations we are using, and at the lowest order 
$L(p,k)= 0$ and therefore $\beta=0$
which implies no symmetry breaking at all. We are forced then to go beyond the
lowest order, the bare graviton,  if we are to find  a non-trivial solution for $\beta $. 
By going to the next leading term, we will
obtain terms which are higher order in $G$ and therefore we will include them only for
$\beta$ and not for $\alpha$ where the bare propagator contribution is the dominant one.

The  standard model possesses a large number of matter degrees of freedom: 12 gauge bosons,
48 chiral fermions (including right handed neutrinos) and 4 Higgs scalars. In extensions
of the standard model the matter content is even larger. This large number of matter
degrees of freedom can help overcome the 
additional $G$ suppression factor arising with the next leading contribution. However,
not all the one loop diagrams enjoy this large $N$ increase.
We will consider only the vacuum polarization diagrams (corrections to the
propagator), and not vertex corrctions, as they are the only ones who benefit
from the large matter content of the theory close to the Planck scale and will exhibit a
large $N$ enhancement.
 
The fact that $L(p,k) = 0$ and therefore the chiral symmetry is unbroken 
at tree level is indeed remarkable. This may be related to gravity's
structure itself, although  one could question whether  this is an 
 artifact of the standard gauge we have chosen.

\begin{figure}[htb]
\begin{minipage}[c][0.35\textheight][t]{0.75\textwidth}
\includegraphics[scale=.95]{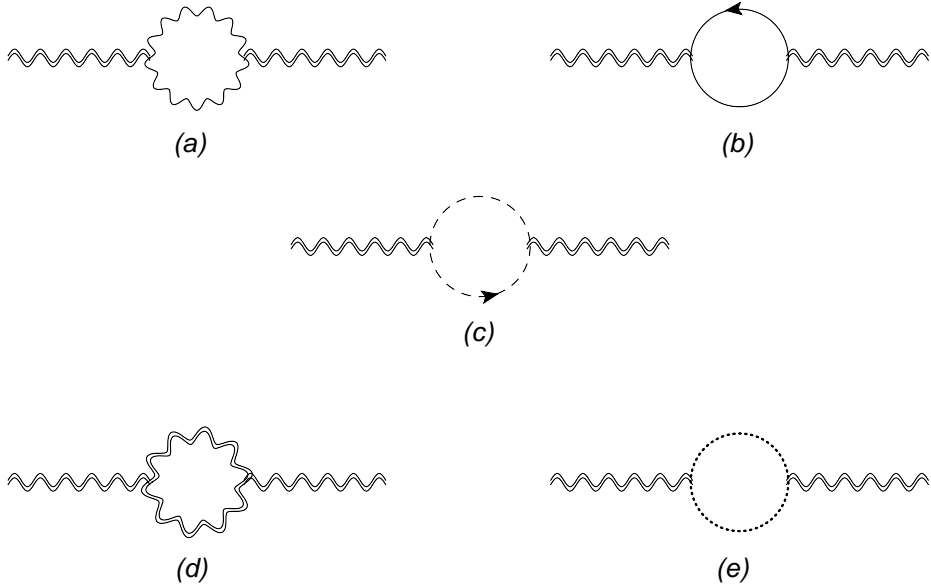}
\end{minipage}
\caption{The set of vacuum polarization diagrams which contribute to the graviton self-energy, which 
comprises gauge bosons (a), Dirac fermions (b), minimal and conformal scalars (c) and gravitons (d).
Notice that there exists a ghost diagram (e) along with the graviton one.}
\label{VacP}
\end{figure}

The matter contribution to the graviton self-energy is obtained from the subdiagrams in Fig.~\ref{VacP} and includes the one loop contributions for gauge bosons, Dirac fermions,
minimal scalars, conformal scalars and gravitons.

The contribution of the photon/gauge boson loop to the polarization tensor reads
\bea
 \tilde\Pi_{\alpha\beta, \gamma\delta} &=& \frac{G}{\pi} \biggl \lbrack \frac{1}{30}\bigl(q_\alpha q_\beta - q^2 \eta_{\alpha\beta} \bigr)\bigl( q_\gamma q_\delta - q^2 \eta_{\gamma \delta}\bigr) 
	 - \frac{1}{20} \bigl( q_\alpha q_\gamma - q^2 \eta_{\alpha\gamma}\bigr) \bigl( q_\beta q_\delta - q^2 \eta_{\beta\delta}\bigr) \notag \\
	&& - \frac{1}{20} \bigl( q_\alpha q_\delta - q^2 \eta_{\alpha\delta}\bigr)\bigl( q_\beta q_\gamma - q^2 \eta_{\beta\gamma}\bigr) \biggr \rbrack \ \log (\mu^2/-q^2) \; ,\label{eq:photon}
\eea
where $q= p-k$ and we have introduced a renormalization mass $\mu $ which is in principle arbitrary.
The gauge boson loop yields as a contribution to the kernel $L(p,k)$ of the $\beta$ term for  the self
energy of the graviton 
\bea
L_{ph}(p,k)= \tau^{\mu\nu} \biggl \lbrack \frac{ \mathcal P_{\mu\nu\alpha\beta}}{q^2}\biggr \rbrack \tilde\Pi^{\alpha\beta, \gamma\delta} (q)\biggl \lbrack \frac{\mathcal
 P_{\gamma\delta \rho \sigma}}{q^2}\biggr \rbrack \tau^{\rho\sigma} \;
	=  -\frac{G^2}{3} \left[ 
(p+k)^2 - \frac{(p^2-k^2)^2}{(p-k)^2}\right]
\log \left(\frac{\mu^2}{-(p-k)^2}\right)
\eea
where $\mathcal  P_{\mu\nu\alpha\beta}$ is the polarization tensor, defined in the appendix.

We have to include also  the Dirac fermion loop (Majorana fermion contribution halves)
\bea
 \hat\Pi_{\alpha\beta, \gamma\delta} &=& \frac{G}{\pi} \biggl \lbrack \frac{1}{60}\bigl(q_\alpha q_\beta - q^2 \eta_{\alpha\beta} \bigr)\bigl( q_\gamma q_\delta - q^2 \eta_{\gamma \delta}\bigr) 
	 - \frac{1}{40} \bigl( q_\alpha q_\gamma - q^2 \eta_{\alpha\gamma}\bigr) \bigl( q_\beta q_\delta - q^2 \eta_{\beta\delta}\bigr) \notag \\
	&& - \frac{1}{40} \bigl( q_\alpha q_\delta - q^2 \eta_{\alpha\delta}\bigr)\bigl( q_\beta q_\gamma - q^2 \eta_{\beta\gamma}\bigr) \biggr \rbrack \ \log (\mu^2/-q^2) \; ,\label{eq:dirac}
\eea
whose addition to the $\beta$ term  input to the self energy
gives
\bea
L_{df}(p,k)=\tau^{\mu\nu} \biggl \lbrack \frac{ \mathcal P_{\mu\nu\alpha\beta}}{q^2}\biggr \rbrack \hat\Pi^{\alpha\beta, \gamma\delta} (q)\biggl \lbrack \frac{\mathcal
 P_{\gamma\delta \rho \sigma}}{q^2}\biggr \rbrack \tau^{\rho\sigma} \;
	=  -\frac{G^2}{6} \left[ 
(p+k)^2 - \frac{(p^2-k^2)^2}{(p-k)^2}\right]
\log \left(\frac{\mu^2}{-(p-k)^2}\right)
\eea
And also minimal scalar fields 
\bea
\bar\Pi_{\alpha\beta, \gamma\delta} &=& \frac{G}{\pi} \biggl \lbrack \frac{1}{40}\bigl(q_\alpha q_\beta - q^2 \eta_{\alpha\beta} \bigr)\bigl( q_\gamma q_\delta - q^2 \eta_{\gamma \delta}\bigr) 
	 - \frac{1}{240} \bigl( q_\alpha q_\gamma - q^2 \eta_{\alpha\gamma}\bigr) \bigl( q_\beta q_\delta - q^2 \eta_{\beta\delta}\bigr) \notag \\
	&& - \frac{1}{240} \bigl( q_\alpha q_\delta - q^2 \eta_{\alpha\delta}\bigr)\bigl( q_\beta q_\gamma - q^2 \eta_{\beta\gamma}\bigr) \biggr \rbrack \ \log (\mu^2/-q^2) \; ,\label{eq:minimalscalar}
\eea
that contribute to the self energy and thus to the kernel of the $\beta $ term by
\bea
L_{ms}(p,k)=\tau^{\mu\nu} \biggl \lbrack \frac{ \mathcal P_{\mu\nu\alpha\beta}}{q^2}\biggr \rbrack \breve\Pi^{\alpha\beta, \gamma\delta} (q)\biggl \lbrack \frac{\mathcal
 P_{\gamma\delta \rho \sigma}}{q^2}\biggr \rbrack \tau^{\rho\sigma} \;
	=  -\frac{3 \; G^2}{8} \left[ 
(p+k)^2 - \frac{2}{3} \frac{(p^2-k^2)^2}{(p-k)^2}\right]
\log \left(\frac{\mu^2}{-(p-k)^2}\right)
\eea
Last but not least we have to include also the contribution of  conformal scalars
\bea
\breve \Pi_{\alpha\beta, \gamma\delta} &=& \frac{G}{\pi} \biggl \lbrack \frac{1}{360}\bigl(q_\alpha q_\beta - q^2 \eta_{\alpha\beta} \bigr)\bigl( q_\gamma q_\delta - q^2 \eta_{\gamma \delta}\bigr) 
	 - \frac{1}{240} \bigl( q_\alpha q_\gamma - q^2 \eta_{\alpha\gamma}\bigr) \bigl( q_\beta q_\delta - q^2 \eta_{\beta\delta}\bigr) \notag \\
	&& - \frac{1}{240} \bigl( q_\alpha q_\delta - q^2 \eta_{\alpha\delta}\bigr)\bigl( q_\beta q_\gamma - q^2 \eta_{\beta\gamma}\bigr) \biggr \rbrack \ \log (\mu^2/-q^2) \; ,\label{eq:conformalscalar}
\eea
whose addition to the kernel of the  $\beta$ function coming from their
contribution to the self energy yields
\bea
L_{cs}(p,k)=\tau^{\mu\nu} \biggl \lbrack \frac{ \mathcal P_{\mu\nu\alpha\beta}}{q^2}\biggr \rbrack  \bar\Pi^{\alpha\beta, \gamma\delta} (q)\biggl \lbrack \frac{\mathcal
 P_{\gamma\delta \rho \sigma}}{q^2}\biggr \rbrack \tau^{\rho\sigma} \;
	=  -\frac{G^2}{36} \left[ 
(p+k)^2 -  \frac{(p^2-k^2)^2}{(p-k)^2}\right]
\log \left(\frac{\mu^2}{-(p-k)^2}\right)
\eea

Despite not being a matter field, the graviton loop contribution to its self energy  
could  prove to be important, although subdominant 
in a formal large $N$  
expansion. Its contribution to the
effective Lagrangian has been worked out by 't Hooft and Veltman \cite{Hoo74}, and from it one can 
estimate its contribution to the vacuum polarization tensor
\bea
\Pi_{\alpha\beta, \gamma\delta} &=& \frac{G}{\pi} \biggl \lbrack \frac{23}{60}\bigl(q_\alpha q_\beta - q^2 \eta_{\alpha\beta} \bigr)\bigl( q_\gamma q_\delta - q^2 \eta_{\gamma \delta}\bigr) 
	 - \frac{7}{40} \bigl( q_\alpha q_\gamma - q^2 \eta_{\alpha\gamma}\bigr) \bigl( q_\beta q_\delta - q^2 \eta_{\beta\delta}\bigr) \notag \\
	&& - \frac{7}{40} \bigl( q_\alpha q_\delta - q^2 \eta_{\alpha\delta}\bigr)\bigl( q_\beta q_\gamma - q^2 \eta_{\beta\gamma}\bigr) \biggr \rbrack \ \log (\mu^2/-q^2) \; ,\label{eq:graviton}
\eea
Inserting this expression into the self energy,  from the graviton loop one obtains the following contribution to the kernel of the $\beta$-term 
\bea
L_{gr}(p,k)=\tau^{\mu\nu} \biggl \lbrack \frac{ \mathcal P_{\mu\nu\alpha\beta}}{q^2}\biggr \rbrack \Pi^{\alpha\beta, \gamma\delta} (q)\biggl \lbrack \frac{\mathcal
 P_{\gamma\delta \rho \sigma}}{q^2}\biggr \rbrack \tau^{\rho\sigma} \;
	=  - 8 G^2 \left[ 
\frac{89}{96} (p+k)^2 -\frac{31}{48} \frac{(p^2-k^2)^2}{(p-k)^2}\right] 
\log \left(\frac{\mu^2}{-(p-k)^2}\right)
\eea

Including all these contributions we obtain 
\bea
\beta(p^2)&=& i \; 2 \pi G^2 \int \frac{d^4 k}{(2\pi)^4} \;\;  \frac{\beta(k^2)}{\alpha(k^2) 
k^2 - \beta^2 k^2} \left[ A (p+k)^2 - B\frac{(p^2-k^2)^2}{(p-k)^2}\right]
\log \left(\frac{\mu^2}{-(p-k)^2}\right) \\
&=& i\; 2 \pi  G^2 \int \frac{d^4 k}{(2\pi)^4} \;\;  \frac{\beta(k^2)}{\alpha(k^2) 
k^2 - \beta^2 k^2} \left\{ A \left[ (p+k)^2 - \frac{(p^2-k^2)^2}{(p-k)^2}\right] + C \frac{(p^2-k^2)^2}{(p-k)^2} \right\}
\log \left(\frac{\mu^2}{-(p-k)^2}\right)
\label{total}
\eea
with
\bea
A&=& -\frac{267+12N_{gb}
+ 6 N_{df} + 27/2 \; N_{ms} + N_{cs}}{288}\\
B&=& \frac{186 +12N_{gb} + 6 N_{df} + 9 
N_{ms} + N_{cs}}{288} \\
C&=&\frac{81  + 9/2  N_{ms}}{288} 
\eea
where $N_{gb}$, $N_{df}$, $N_{ms}$ and $N_{cs}$ correspond to the number of
gauge bosons, Dirac fermions, minimal scalar and conformal scalar degrees of
freeedom in the model, respectively.

In eq.(\ref{total}) the conformal contributions are contained in the $A$ term
while conformal breaking terms contribute to the $C$ term. While both terms
can help trigger dynamical symmetry breaking, the $A$ term for small
external moment, $p \ll k$, becomes
\bea
\left[ (p+k)^2 - \frac{(p^2-k^2)^2}{(p-k)^2}\right] \xrightarrow{p \ll k} - (k \cdot p)^2 +
 4 k^2 p^2 \longrightarrow 0
\eea
so that it cannot give rise to an infrared mass.
Thus, it is only the term proportional to $C$  that can provide a dynamical infrared mass
to the right handed neutrino,
effectively breaking both symmetries, lepton number and chiral symmetry, dynamically.  
The infrared mass generated will be supressed with respect to the cut-off scale
by the factor $C/A$ which is $\mathcal O (1/N_{\mbox{matter}})$.

Now imposing  $\alpha$  positive and $\beta$ real for space-like momentum,
which is imposing that the fermion is stable without
 a tachyonic singularity, we can safely perform the angular integration
after the Wick rotation. We get
\bea
\alpha (x) = 1 - \frac{G \; \Lambda^2}{32 \pi} \int_0^1 dy \; K(x,y) 
\frac{y \; \alpha(y)}{ y \alpha^2(y) + \beta^2(y)}
\label{eq:alpha}
\eea
and
\bea
\beta (x) =  \frac{G^2 \; \Lambda^4}{16 \pi} \int_0^1 dy \; L(x,y) 
\frac{y \; \beta(y)}{ y \alpha^2(y) + \beta^2(y)}
\label{eq:L}
\eea
where $x= p^2/\Lambda^2$ and $y= k^2/\Lambda^2$ and
\bea
K(x,y)& =&  \frac{1}{x} \left[ \frac{16 (x^2+y^2) + 64 xy}{x+y+|x-y|} -8 (x+y) \right]
\nonumber\\
&&\nonumber \\
L(x,y)& = &\frac{A}{48 x y} \left\{ -2 |x-y| \left(5 x^2 + 2 x y + 5 y^2 \right) + (x+y)
\left(-36 xy + 10 (x+y)^2 \right) + 24 xy (x+y) \log\left[\frac{x+y+|x-y|}{2}\right]
\right\} \nonumber \\
&& \nonumber \\
&&+ \frac{B}{8 x y}\left\{ 2|x-y| \left[ 1 -2 \log\left[\frac{2 (x-y)^2}{x+y+|x-y|}\right]
\right]+
(x+y) \left[ -2 + \log\left[ \frac{-2 x y + (x+y)(x+y+|x-y|)}{2}\right]\right]\right\}
\nonumber
\eea
We have introduced an UV momentum cut-off $\Lambda $, arising from the non-renormalizabilty
of the theory which does not necessarily possess the same value as the scale $\mu$. However
one would expect them to be of the same order of magnitude. In the remaining of this
work we will consider them to be equal, $\Lambda = \mu = 0.8 M_{Pl}$.

As in \cite{Abe:1983in}, we implement 
the following iterative method to find the numerical solutions of the
Schwinger-Dyson equations. We first choose two trial functions to start our
iterative calculation
\bea
\alpha^{(0)} (x) = c_1 \;\;\; , \beta^{(0)} (x) = c_2
\eea
with $c_1$ and $c_2$ constants. We then define $\alpha^{(i)}$ and $\beta^{(i)}$ by
\bea
\alpha^{(i+1)} (x) = 1 - \frac{G \; \Lambda^2}{32 \pi} \int_0^1 dy \; K(x,y) 
\frac{y \; \alpha^{(i)}(y)}{ y \alpha^{(i) \;2 }(y) + \beta^{(i)\; 2 }(y)}
\eea
and
\bea
\beta^{(i+1)} (x) = \frac{G^2 \; \Lambda^4}{16 \pi} \int_0^1 dy \; L(x,y) 
\frac{y \; \beta^{(i)}(y)}{ y \alpha^{(i) \; 2}(y) + \beta^{(i) \;2}(y)}
\eea
expecting that the series ${\alpha^{(i)}(x)}$ and ${\beta^{(i)}(x)}$ converge into the 
solutionts of the Schwinger-Dyson equations $\alpha(x) $ and $\beta(x)$, respectively.
In practice, we assume that the series ${\alpha^{(i)}(x)}$ and ${\beta^{(i)}(x)}$ converge
if the normalized difference $\mid \alpha^{(i+1)} - \alpha^{(i)} \mid + 
\mid \beta^{(i+1)} - \beta^{(i)}\mid $ become less than $10^{-3}$.

For a true dynamical symmetry breaking  one 
has to check that the minimum we have found is a real minimum. For that 
purpose, we use the effective potential defined in \cite{Cornwall:1974vz}. 
In the Hartree-Fock approximation the effective potential $V[S]$
as the functional of the fermion propagator $S$ is obtained to be
\bea
V[S] &=& -i \int \frac{d^4 p}{(2 \pi)^4} \mbox{Tr} \left[ \log S_F^{-1}(p) \; S(p) - 
S_F^{-1}(p) \; S(p) +1 \right] \nonumber \\
& & -\frac{i}{2} \int \frac{d^4 p \; d^4 k}{(2 \pi)^8} \mbox{Tr} \left[ \Gamma_{\alpha \beta}
(k,p-k) \; S(p) \; \Gamma_{\mu \nu} (p,k-p) \; S(k) \; G^{\alpha \beta \mu \nu}(k-p) \right]
\eea
Substituting the solution of the Swinger-Dyson equation 
$\delta V[S_F^\prime]/\delta S_F^\prime=0$ into the
above expression we get
\bea
V[S_F] &=& \frac{-i}{2} \int \frac{d^4 p}{(2 \pi)^4} \mbox{Tr} \left[ 2 \log S_F^{-1}(p) \; S_F^\prime(p) -  S_F^{-1}(p) \; S_F^\prime(p) +1 \right] 
\eea
Using now that $S_F(p)= i/\slashed p $ and eq(\ref{eq:fermionprop}) and performing the
angular integration after Wick's rotation we have
\bea
V[\alpha ,\beta] = \frac{\Lambda^4}{8 \pi}^2 \int_0^1 dx \left\{ \log \frac{x}{x \alpha^2(x)
+ \beta^2(x)} - \frac{x \alpha(x)}{x \alpha^2(x) + \beta^2(x)} +1 \right\}
\eea
where $\alpha$ and $\beta$ are the solutions of eqs (\ref{eq:alpha})-(\ref{eq:L}).
As the solution corresponding to the true vacuum should minimize the potential, if there
is a symmetry breaking solution ($\alpha= \alpha_\beta , \beta \neq 0$) besides the symmetry
conserving solution ($\alpha = \alpha_o , \beta=0 $) a difference of potentials
\bea
\delta V = V[\alpha= \alpha_\beta , \beta \neq 0] - 
V[\alpha = \alpha_o , \beta=0 ]
\eea
selects the solution of the two corresponding to a true vacuum.

\begin{figure}[htb]
\begin{minipage}[c][0.30\textheight][t]{0.45\textwidth}
\includegraphics[scale=.95]{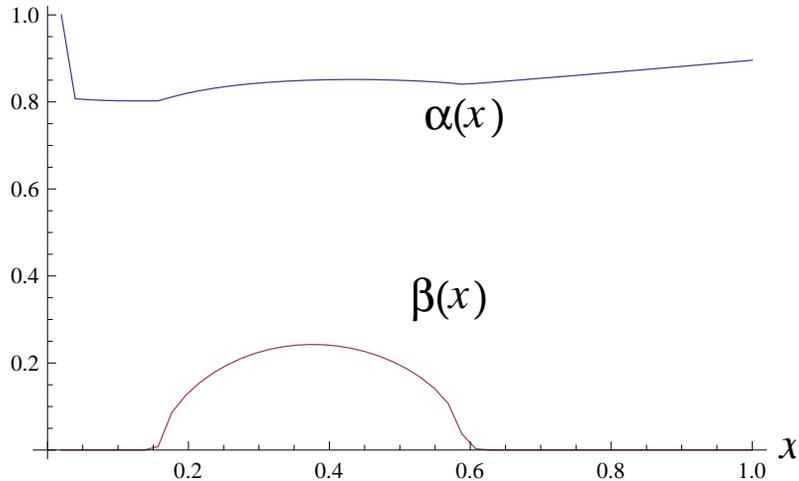}
\end{minipage}
\caption{The symmetry breaking solutions of the Schwinger-Dyson equation, $\alpha$ and
$\beta$ for $A=10$  as a function of $x=p^2/\Lambda^2$}
\label{alphabeta}
\end{figure}

In the standard model $A=203/96 \approx 2.1$ which is not large enough to
produce a symmetry breaking minimum. A much larger matter content would be needed
such that $A>8$ for that purpose. 
Cut-off scales even closer to $M_{Pl}$ would allow symmetry breaking for
smaller $A$. In minimal supersymmetry, with  12 gauge bosons,
80 chiral fermions and 56 real (minimal) scalars,
$A$ turns out to be $A \approx 5$.
The particle content again is not enough to produce symmetry breaking. However,
extended models or models with extra dmensions, can reach values of $A$ in the desired range.
For example in N=8 supergravity \cite{Lykken:1996xt}
 $A>8$.
In Fig \ref{alphabeta} we show the
symmetry breaking solution of the Schwinger-Dyson equation for $A=10$.

\begin{figure}[htb]
\begin{minipage}[c][0.30\textheight][t]{0.45\textwidth}
\includegraphics[scale=.95]{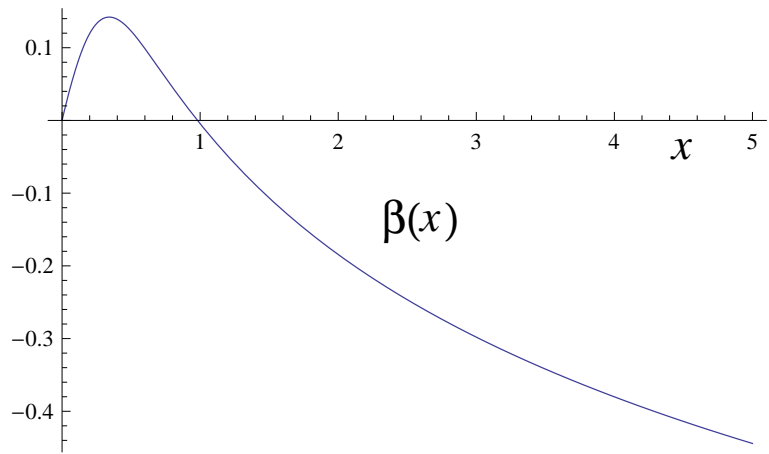}
\end{minipage}
\caption{The solution of the Schwinger-Dyson equation $\beta(x)$ becomes repulsive for momenta
larger than the cut-off scale $\Lambda $}
\label{beta}
\end{figure}

Fig \ref{beta} shows our solution for $\beta (x)$ versus  $x=p^2/\Lambda^2$ for values of
momenta larger than the cut-off scale. It is evident there that the 
cut-off signals when the interaction stops being attactive. In this sense, its existence
seems justified.
\begin{figure}[htb]
\begin{minipage}[c][0.30\textheight][t]{0.45\textwidth}
\includegraphics[scale=.95]{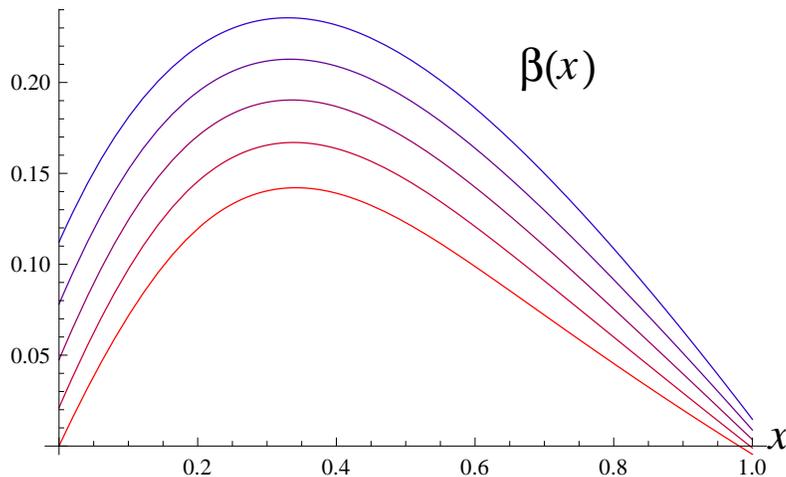}
\end{minipage}
\caption{$\beta(x)$ for $C/A = 0.2 \mbox{(red)},0.4,0.6,0.8,1\mbox{(blue)}$ as a function of
$x = p^2/\Lambda^2$ 
The larger the ratio, the heavier
the infrared mass}
\label{infraredmass}
\end{figure}
Once the symmetry is broken a mass for the right handed neutrino will be produced. Such a
mass, arises from the conformal breaking part of the graviton or the minimal scalar loop contributions  (the $C$ term), 
and  is roughly 
given by 
\bea
m_{RH} \approx  \frac{C}{A}  \; \frac{\Lambda}{4 \pi} \;\sim \; \mathcal O(1/N) .
\eea
where $N$ reflects matter field contribution to the conformal term.
For the successful symmetry breaking scenario shown before,
with $A =10$, it produces a mass $m_{RH} \approx 10^{-3} \Lambda$, 
exactly in the right ball-park to a succesful see-saw.

\section{Resummation}

\label{sec:res}

As we have seen in the previous section, the large matter content of the standard model
(and the even larger one of extended theories) could help overcome the $G$ suppression of the
next leading order contribution to the kernel of the $\beta $ function, opening the
door to a dynamical symmetry breaking. If this is the case, then some additional, higher order 
contributions, suppressed, in principle by extra powers of $G$ would also benefit
from the $N$ enhancement and must be included in the calculation. 

Following Tomboulis \cite{Tomboulis:1977jk}
 we can include those contributions. As $N$ (the number of matter fields)  
becomes  large  (even if not as large and we 
would have needed in the previous section), a natural expansion
would be  then an expansion in powers of $1/N$, keeping $G N$ fixed. 
Hence, we will explore the $1/N$ expansion where some
systematic resumation of the perturbative series can be done.

The general validity of this $1/N$ expansion is based on the fact that gravity
couples universally to all matter. Besides, the expansion possesses important
advantages which make it very convenient to use. Unlike the coupling $G$, $N$ is
a scale independent parameter. Therefore, the solution to the theory to a given
order in $1/N$ as given by the expansion holds in principle for all momenta scales,
and can be used to probe the strong coupling regime. Furthermore, being $N$  a gauge
invariant parameter, the expansion is manifestly gauge invariant.

To proceed with the expansion, notice that to leading order in $1/N$ 
we should include only those graphs which contain  closed matter loops  including
fermions, gauge bosons or conformaly coupled scalars.
The large number of  
matter fields defines a large $N$ enhancement that compensates for the  
Planck suppression of the additional coupling factor $G$ producing
an effective ``reduced" Planck mass,
$M_{Pl \; \mbox{\small{reduced}}}^2 =  M_{Pl}^2/N $.
In fact the enhancement occurs only for the spin 2 part of the propagator, the $A$ term
and not the trace part  
as reflected in the $C$ term. It is not necessary to include the
vertex corrections as they enter only at the next level in the $1/N$ expansion.
\begin{figure}[htb]
\begin{minipage}[c][0.07\textheight][t]{0.75\textwidth}
\includegraphics[scale=.95]{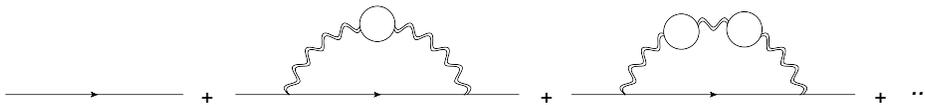}
\end{minipage}
\caption{Only those graphs which contain a closed loop of matter fields benefit from the
large $N$ enhancement and are to be resummed}
\label{resum}
\end{figure}

\begin{figure}[htb]
\begin{minipage}[c][0.30\textheight][t]{0.45\textwidth}
\includegraphics[scale=.95]{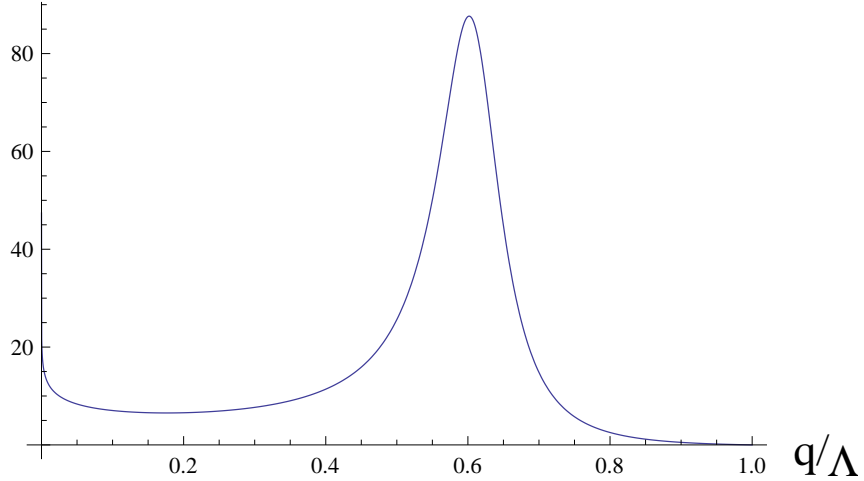}
\end{minipage}
\caption{Large $N$ enhancement of the space-like propagator in arbitrary units as a function of the momentum trasfer
(in units of the cut-off $\Lambda$) }
\label{prop}
\end{figure}

Therefore, the matter, gauge boson and
conformally coupled scalar loop corrections to the bare 
graviton propagator must now be included as part of the leading order graviton propagator.
Using the  one loop contribution of such particles to the graviton self energy
calculated before we obtain for the complete leading order graviton propagator
\bea
D_{\mu \nu \alpha \beta}= \frac{ \mathcal P^{(2)}_{\mu\nu\alpha\beta}}{q^2 \left( 1 - \frac{2}{ \pi } N G q^2 
\log \left(\frac{-q^2}{\mu^2 }\right)  \right)}
\eea
where $\mathcal P^{(2)} $ is the spin 2 part of the propagator and 
$N$ is a measure of the number of matter fields in the theory, $N   \propto N_{gb} + N_{df}/2 +N_{(cs+ms)}/12 $. In Fig \ref{prop} the potential enhancement of the
spacelike propagator 
due to large $N$ is depicted. Such enhancement is due to the fact that 
the leading $1/N$ propagator posseses no spacelike spin 2 poles for real momenta. Instead,
it posseses pairs of complex conjugate poles in the complex plane.
As shown by Tomboulis \cite{Tomboulis:1977jk,Tomboulis:1980bs}, 
the gravitational quantum
corrections modify the propagator in such a way that the new propagator has no
unphysical poles on the real $q^2$ axis. The would-be unphysical poles
are shifted off the real axis by the matter interactions and split into a pair of
complex conjugate poles, i.e.  quantum corrections shift the tachyonic ghost poles
 to a pair of
complex poles on the physical sheet. 
According to Lee and Wick \cite{Lee:1969fy} the theory can be defined so that
the complex poles do not contribute to the 
absorptive part of the amplitudes and
physical unitarity can be maintained. However, the standard analyticity
properties of the $S$ matrix are modified by the extra complex poles which
results in a breakdown of locality. This should come as no surprise as we expect
gravity to be non-local once quantum effects become important.

With this interpretation we can proceed with the resummation. We begin by re-writting the 
different contributions to the polarization tensor as a traceless spin 2 and a spin 0 part
\bea
\Pi_{\alpha\beta, \gamma\delta} &=&  \frac{\Pi_2}{q^4} \left[ \bigl( q_\alpha q_\gamma - q^2 \eta_{\alpha\gamma}\bigr) \bigl( q_\beta q_\delta - q^2 \eta_{\beta\delta}\bigr) +
\bigl( q_\alpha q_\delta - q^2 \eta_{\alpha\delta}\bigr)\bigl( q_\beta q_\gamma - q^2 \eta_{\beta\gamma}\bigr) - 
\right. \nonumber \\
&& \left. \frac{2}{3} 
(q_\alpha q_\beta - q^2 \eta_{\alpha\beta} \bigr)\bigl( q_\gamma q_\delta - q^2 \eta_{\gamma \delta}\bigr) \right]_2 + 
\frac{\Pi_o}{q^4}
 \left[(q_\alpha q_\beta - q^2 \eta_{\alpha\beta} \bigr)\bigl( q_\gamma q_\delta - q^2 \eta_{\gamma \delta}\bigr)\right]_0 
\eea
define
\bea
\hat \eta_{\mu \nu} \equiv q_\mu q_\nu - q^2 \eta_{\mu \nu}
\eea
and sum all the bubbles
\bea
D_{\mu \nu \alpha \beta} &=& \frac{P}{q^2} + \frac{P}{q^2} \Pi \frac{P}{q^2} +
\frac{P}{q^2} \Pi \frac{P}{q^2} \Pi \frac{P}{q^2} + 
\frac{P}{q^2} \Pi \frac{P}{q^2} \Pi \frac{P}{q^2} \Pi \frac{P}{q^2} 
+ \dots
\nonumber \\
&&\nonumber \\
&& 
\frac{P}{q^2}  + \frac{\Pi_2}{(q^2)^2}\frac{1}{1 -2 \frac{\Pi_2}{q^2}} \left[
\hat\eta_{\mu \alpha} \hat \eta_{\nu \beta} +  \hat\eta_{\nu \alpha} \hat \eta_{\mu \beta} 
- \frac{2}{3} \hat\eta_{ \alpha \beta } \hat \eta_{\mu \nu} \right] \nonumber + \frac{\Pi_0}{(q^2)^2}\frac{1}{1 + \frac{3}{2} \frac{\Pi_0}{q^2}} \left[
\left(\hat \eta_{\mu \nu} - \frac{3}{2} \eta_{\mu \nu} \right) \left(
\hat \eta_{\alpha \beta} - \frac{3}{2} \eta_{\alpha \beta} \right) \right]
\eea  
where $P$ is the graviton polarization sum $ \mathcal P^{\mu\nu\alpha\beta} $ 
in the harmonic gauge and 
$\Pi$ represents the polarization tensor (Lorentz indices have been omitted).
Then the resummed kernel becomes
\bea
L_{rs}(p,k)=\tau_{\mu \nu} D^{\mu \nu \alpha \beta} \tau_{\alpha \beta} = 
\frac{8 \pi G}{(q^2)^2} \left\{ \frac{\Pi_2}{1 -2 \frac{\Pi_2}{q^2}} \frac{5}{6} \left[
(p+k)^2 - \frac{(p^2-k^2)^2}{(p-k)^2}\right] + 
\frac{\Pi_0}{1 + \frac{3}{2} \frac{\Pi_0}{q^2}} \left[\frac{25}{16}
(p+k)^2 - \frac{(p^2-k^2)^2}{(p-k)^2}\right] \right\}
\nonumber
\eea
with
\bea
\Pi_2(q^2) = -\frac{G}{\pi} (q^2)^2 \log\left( \frac{\mu^2}{-q^2}\right) \left[
\frac{7}{40} + \frac{1}{20} N_{gb} + \frac{1}{40} N_{df} + \frac{1}{240} N_{ms} +
\frac{1}{240} N_{cs} \right]
\eea
and
\bea
\Pi_0(q^2)= -\frac{G}{\pi} (q^2)^2 \log\left( \frac{\mu^2}{-q^2}\right) \left[
\frac{1}{2} + \frac{1}{36} N_{ms} \right]
\eea
with this the spin 2 denominator takes the form
\bea
1 -2 \frac{\Pi_2 (q^2)}{q^2} = 1 + \frac{2 G}{\pi} q^2 \log\left( \frac{\mu^2}{-q^2}\right) \left[
\frac{7}{40} + \frac{1}{20} N_{gb} + \frac{1}{40} N_{df} + \frac{1}{240} N_{ms} +
\frac{1}{240} N_{cs} \right]
\eea
while that of the spin 0 denominator gives
\bea
1 + \frac{3}{2} \frac{\Pi_0}{q^2} = 1 -\frac{3}{2}\frac{2 G}{\pi} q^2  \log\left( \frac{\mu^2}{-q^2}\right) \left[
\frac{1}{2} + \frac{1}{36} N_{ms} \right]
\eea
notice that although both denominators
 go like $ \frac{q^2}{\pi}  \log\left( \frac{\mu^2}{q^2}\right) $, only the spin 2 
part, the $A$ term,  has the large $N$ enhancement.
The large enhancement shown in Fig \ref{prop} occurs when the complex poles are close
to the real axis as would be expected near criticality in the Lee-Wick theory.
Thus enhancement is achieved by tuning the reduced Planck scale, i.e. by tuning the
matter content of the theory and  keeping the renormalization scale $\mu$ always below but
not far from its 
maximum possible
value $\mu_{\mbox{max}} = \sqrt{e}  M_{Pl \; \mbox{\small{reduced}}}$.

\begin{figure}[htb]
\begin{minipage}[c][0.42\textheight][t]{0.85\textwidth}
\includegraphics[scale=.5]{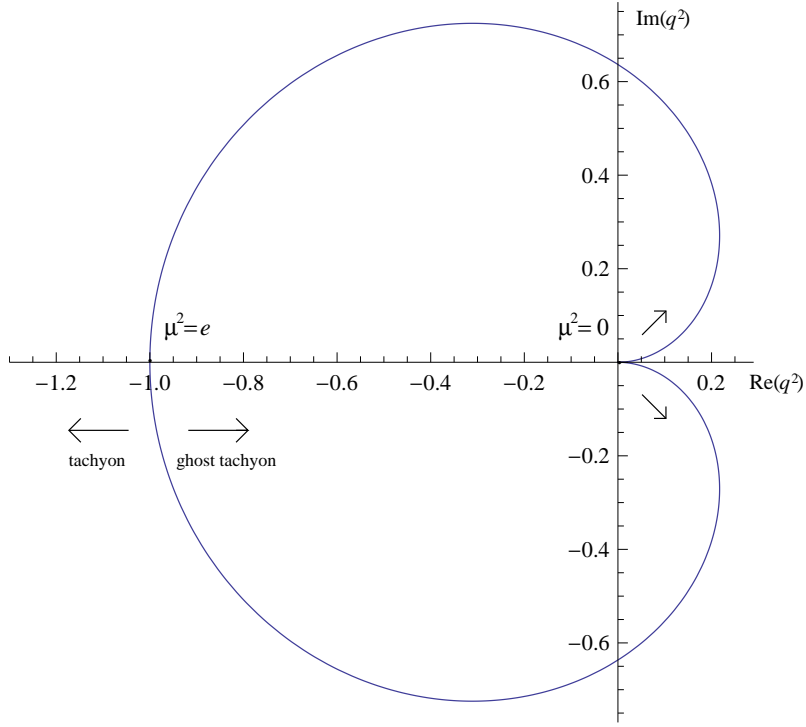}
\end{minipage}
\caption{The trajectory of the poles in the complex plane is shown as a function of the scale $\mu $
in units of the reduced Planck mass. }
\label{poles}
\end{figure}

In Fig \ref{poles} we show how the poles travel the
complex plane as a function of the renormalizaton scale $\mu$.
The spin 0 part, the $C$
term, even if close to the phase transition point does not get enhanced in the large $N$
scenario.  The vast increase displayed in Fig \ref{prop} is attained when the maximum of
the  $ q^2 \log\left( \frac{\mu^2}{q^2}\right) $ term in the denominator of the $A$ term
is tuned to be close to 1.

\begin{figure}[htb]
\begin{minipage}[c][0.30\textheight][t]{0.45\textwidth}
\includegraphics[scale=.95]{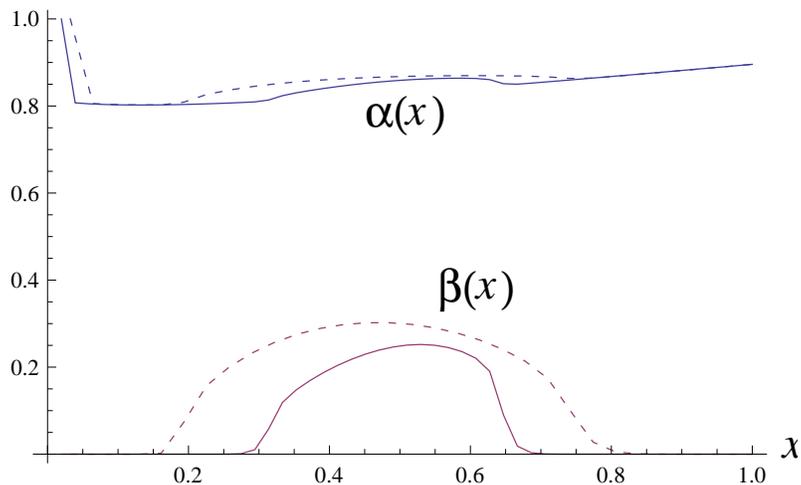}
\end{minipage}
\caption{The symmetry breaking solutions of the Schwinger-Dyson equation, $\alpha$ and
$\beta$ after resummation  as a function of $x=p^2/\Lambda^2$ for a matter content consistent
with supersymmetry using the narrow width approximation (solid line) and numerically integrating the
exact expression (dashed line)}
\label{alphabetars}
\end{figure}

One can then expand the $ \frac{q^2}{\pi}  \log\left( \frac{\mu^2}{q^2}\right) $
term around its maximum, $Q^2 = \mu^2/e $, as in the narrow width approximation, i.e.
we can approximate the propagator shown before by a gaussian,
so that we can estimate the size (the strength) of the spin 2 contribution, while
neglecting that of the spin 0 part. 
For the spin-2 denominator we  get
\bea
1 -2 \frac{\Pi_2 (q^2)}{q^2} = && 1 - \frac{2}{\pi} G \frac{\mu^2}{e} \left[
\frac{7}{40} + \frac{1}{20} N_{gb} + \frac{1}{40} N_{df} + \frac{1}{240} N_{ms} +
\frac{1}{240} N_{cs} \right]  + \nonumber \\
&&
\frac{1}{\pi} G \frac{e}{\mu^2} \left[
\frac{7}{40} + \frac{1}{20} N_{gb} + \frac{1}{40} N_{df} + \frac{1}{240} N_{ms} +
\frac{1}{240} N_{cs} \right] \left(Q^2 -\frac{\mu^2}{e}\right)^2  + \dots 
\eea
which under the narrow width approximation becomes
\bea
&& 1 - \frac{2}{\pi} G \frac{\mu^2}{e} \left[
\frac{7}{40} + \frac{1}{20} N_{gb} + \frac{1}{40} N_{df} + \frac{1}{240} N_{ms} +
\frac{1}{240} N_{cs} \right]  = \nonumber \\
&&\frac{1}{\pi} G \frac{e}{\mu^2} \left[
\frac{7}{40} + \frac{1}{20} N_{gb} + \frac{1}{40} N_{df} + \frac{1}{240} N_{ms} +
\frac{1}{240} N_{cs} \right] \Delta^2 \ll 1
\eea
so that
\bea
L_{rs}(p,k)= \tau_{\mu \nu} D^{\mu \nu \alpha \beta} \tau_{\alpha \beta} \longrightarrow
\frac{20 \pi}{3} G \frac{\mu^2}{e} \frac{ 2 p^2 + 2 k^2 - \mu^2/e - (p^2 -k^2)^2 e/\mu^2}
{\Delta^2 + (Q^2 -\mu^2/e)^2}
\eea
thus
\bea
\beta(p^2)&=&  \int \frac{d^4 k}{(2\pi)^4} \;\;  \frac{\beta(k^2)}{\alpha(k^2) 
k^2 - \beta^2 k^2} \; \; \tau_{\mu \nu} D^{\mu \nu \alpha \beta} \tau_{\alpha \beta}
\Biggl|_{(p-k)^2=Q^2} \Biggr.
\eea
One can then perform the angular integration after the Wick rotation to find
\bea
\beta (x) &= & \frac{G^2 \; \Lambda^2 \mu^2 }{16 \pi \;e } \int_0^1 dy \; L_{rs}(x,y) 
\frac{y \; \beta(y)}{ y \alpha^2(y) + \beta^2(y)} \nonumber
\eea
\bea
L_{rs}(x,y)&= & \frac{5}{12}  \left[ \Delta - \sqrt[4]{\left( \Delta^2 + \left( -R^2 -2 \sqrt{x y} +x +y \right)^2
\right) \left( \Delta^2 + \left( -R^2 +2 \sqrt{x y} +x +y \right)^2
\right) }\right]  \nonumber \\
 && \left(\frac{ 2 x + 2 y - R^2 - (x-y)^2/R^2  }{ 4 x y \Delta} \right) \;\;
\cos \left[ \frac{\mbox{Arg}\left(  2 x \left(- i \Delta + R^2 + y \right) + 
\left( \Delta + i (R^2-y)^2 \right)^2 -x^2 \right)}{2} \right] 
\eea
where $R^2  = \mu^2/e$.
Now, for a much smaller $N$, we can achieve symmetry breaking. Specifically,
symmetry breaking is accomplished for 
\bea
\frac{7}{40} + \frac{1}{20} N_{gb} + \frac{1}{40} N_{df} + \frac{1}{240} N_{ms} +
\frac{1}{240} N_{cs} > 3
\eea
In Fig. \ref{alphabetars} the solutions of the Schwinger-Dyson equation
are shown for the resummed propagator under the narrow width approximation using an $N$ 
consistent with the matter content of supersymmetry (solid line).
Despite being useful, as it allows to perform analitically the angular integration,
the narrow width approximation does not capture the low momentum behaviour of the
propagator and therefore offers a conservative estimate of the minimum matter content
necessary to achieve dynamical symmetry breaking. A more accurate result can be
obtained by numerically integrating the exact expression. The result is depicted
in Fig. \ref{alphabetars} in dashed line. In this case, slightly smaller matter
contents also admit dynamical symmetry breaking.

\section{Pure large $N$}

As we have seen the Schwinger-Dyson equations, which contain 
the full dynamical information of the quantum field gauge
theory can serve as an adequate and effective tool for a non-perturbative
approach to gravity.  However, they consist on an infinite chain
of strongly coupled highly nonlinear integral equations,
and some truncation scheme is needed in order to make these
equations tractable for extracting physical information from them.
At leading order,  it is natural to assume that approximating the full vertices by their free perturbative (point-like) counterparts (and in addition replacing the full 
graviton propagator  by its free perturbative expression) in the corresponding kernels 
of the above mentioned integral equations is consistent. 
Nonetheless it is known that such an approximation does not 
preserve explicit gravitational gauge covariance (general coordinate invariance).

Beyond the leading order, the matter contributions are separately 
gauge invariant at one loop and at large $N$.  This is not clear for the 
gravitational loop contributions. There is a similar problem in QCD 
where the gluon loop contributions are partly canceled by vertex 
corrections. In this case it seems possible to take into account self-interacting gluon modes
(the non-Abelian character of QCD) at the level of the full gluon propagator only (the so
called improved ladder approximation \cite{ila}), hence vertices
remain intact, i.e. bare ones.

We will not follow this avenue here, instead we would like to consider the pure large $N$ 
regime, i.e. to consider in  both the one loop calculation and the resummation {\bf only } those contributions that exhibit the large $N$ enhancement and that are gauge invariant
by themselves. Therefore we will eliminate the graviton loop contribution to the
graviton self energy and assume that vertex or other contributions 
to the Schwinger-Dyson 
equations cancel the 't Hoof-Veltman contribution.

As we have seen, neither in the one loop nor in the resummed case, the graviton
loop share to the kernel of the $\beta $ function is particularly important,
therefore we do not expect our results regarding symmetry breaking to change in a significant way.
However, the graviton loop was the key (together with the minimal scalar loop)
to generate a conformal breaking term. Such a term controls the existence
(and the magnitude) of the dynamically generated  infrared mass. Then question is then, whether by
suppressing the graviton loop, the infrared mass will disappear as well.
The aswer to this question is no. 
The infrared mass  could  be generated by the known conformal anomalies of the matter 
fields - the trace anomalies.   These would be suppressed by 
additional loop and coupling constant effects and will end up being roughly in the
right ball-park to generate realistic right-handed neutrino  
masses.

We can now analyze the impact of the suppression of the graviton loop in the
results obtained before. In the one loop calculation of the graviton propagator
(in the large $N$ regime) we have seen that dynamical symmetry breaking is achieved
for $A >8$. This remains being correct,  except that (in the absence of minimal scalars)
$A$ is now given by
\bea
A&=& -\frac{12N_{gb}
+ 6 N_{df}  + N_{cs}}{288}
\eea
and $C =0$.  As a result, more matter fields are needed to compensate the missing
contribution. Roughly 12 additional gauge bosons or 24 additional fermions.
 
\begin{figure}[htb]
\begin{minipage}[c][0.30\textheight][t]{0.45\textwidth}
\includegraphics[scale=.95]{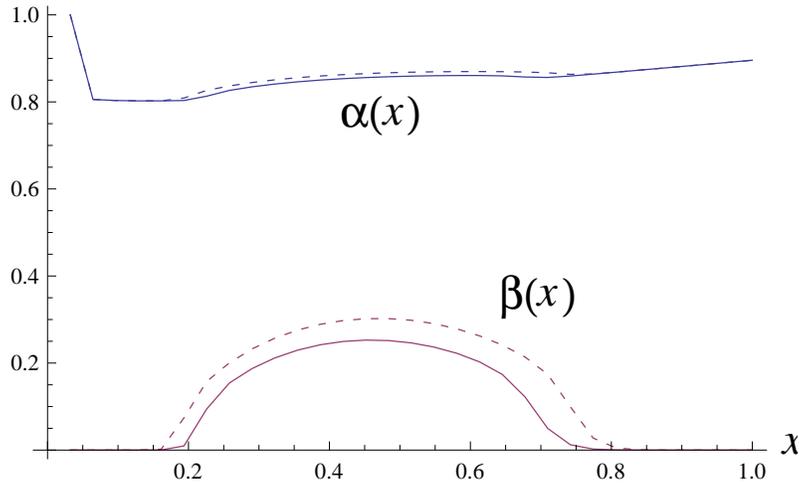}
\end{minipage}
\caption{The exact symmetry breaking solutions of the Schwinger-Dyson equation, $\alpha$ and
$\beta$ after resummation  as a function of $x=p^2/\Lambda^2$ for a matter content consistent
with supersymmetry with  (dashed line) and without (solid line)the graviton loop input}
\label{alphabetaln}
\end{figure}

In the resummed case, (in the absence of minimal scalars) 
the condition for symmetry breaking now reads
\bea
 \frac{1}{20} N_{gb} + \frac{1}{40} N_{df} +
\frac{1}{240} N_{cs} > 3
\eea
In this case just 7 additional fermions, or 4 additional gauge bosons will
offset the graviton loop lost.
In Fig. \ref{alphabetaln} the exact solutions of the Schwinger-Dyson equation
are shown for the  resummed propagator  using an $N$ 
consistent with the matter content of supersymmetry in the pure large $N$ regime (solid line).
For comparison, the solutions including the graviton loop contribution to the graviton
self energy are also shown (dashed line).

\section{Discussion and Conclusions}

\label{sec:con}
In this work we studied the possibility that an enhanced gravitational
attraction can spark the formation of a right handed neutrino
condensate, inducing dynamical symmetry breaking and generating a Majorana
mass for the right handed neutrino at a scale appropriate for the see-saw 
mechanism. The composite field formed by the condensate phase could drive an early epoch
of inflation leaving an imprint tightly constrained, that can be experimentally
probed in the near future.

The major task in studying chiral symmetry breaking is to establish the
Schwinger-Dyson equation for the fermion self-energy, which takes into
account the non-perturbative features of the theory and then, to investigate
whether this equation admits a non-zero fermion mass as a solution.

We have found that to the lowest order, the theory does not allow dynamical
symmetry breaking.
Nevertheless, thanks to the large number of matter fields
in the standard model (and the ever larger one in extended models) the 
 suppression by an additional power in $G$ in the next leading term can be
compensated, enhancing these higher order terms up to their lowest order counterparts. 
We have seen then, that for a vast number of matter
fields chiral symmetry can be broken dynamically and the infrared mass
generated this way turns out to be in the expected range for a successful see-saw scenario.

To further exploit the large $N$ potential, we considered a $1/N$ expansion of
the graviton propagator, keeping $G N$ fixed. The resulting lowest order graviton
propagator has $1/(q^4 \log(q^2))$ asymptotic behaviour and no tachyon or unphysical
real bound state poles. There are however, complex conjugate poles on the physical
sheet that when close to the real axis, significatively enhance the spacelike
propagator. This enhancement allows a dynamical symmetry breaking solution for a matter
content consistent with the standard model.

A distinguishing feature of the model is the presence of non-local effects
associated with the complex poles. Such effects become appreciable only
at ultra high energy scales. It will be quite interesting to investigate
the effects of these nonlocalities in the very early universe where they
become relevant, in particular with respect to the horizon and homogeneity problems.

Probably the least attractive feature of our model is the fact that we are working
at energies close to Planck scale, where our physics knowledge is rather poor.
It is widely believed  that field theory breaks
down at  Planck scale because gravity becomes
important and a full theory of quantum gravity is called
for.  If the gauge force strength is characterized by the dimensionless gravitational
coupling  $\alpha_G= G^2 \Lambda^2  =\Lambda^2/M_{Pl}^2$
at an energy scale $\Lambda$, then
yet unknown quantum gravity effects become dominant at energy scales close to $M_{Pl}$
 where $\alpha_G$  approaches 1, the regime needed for dynamical symmetry breaking.
However, since gravity (and thus
the graviton) couples to all fields, matter loop corrections can enhance 
the effective  gravitational coupling to the needed range, 
before full quantum gravity completely blurs the picture.

In our case,  it is crucial to notice that 
there are (at least) three relevant scales, one fundamental scale, the
original Planck scale, and two dynamically induced scales, the reduced Planck scale, 
$M_{Pl \; \mbox{\small{reduced}}}^2 =  M_{Pl}^2/N $,  that governs size of 
the net matter loop corrections and the scale where the dominant chiral 
symmetry breaking occurs for the condensate.  
Of course, there is also the scale of the infrared mass 
which is suppressed relative to the others scales and is relevant for 
the phenomenology of the right-handed neutrinos. The reason for this 
suppression is that the symmetry breaking is largely due to the $A$ term,
which is the dominant contribution to the kernel for the
 $\beta $ function, while 
the infrared mass is generated by the conformal breaking $C$ term, which 
does not enjoy the large $N$ enhancement.

As we have seen the  dynamical scales are 
suppressed relative the the original Planck scale courtesy of the
large matter content of the model, which implies an
inevitable hierarchy between the energy scale of the field
theory and the Planck scale. 
This effectively amounts to an enhancemnet
of the dimensionless gravitational coupling, $\alpha_{G \; \mbox{\small{large $N$}}} =  
N \alpha_{G} $ at momenta that are below the normal Planck scale opening the
door to dynamical symmetry breaking at those scales.
We assume that as the energies involved  are sufficiently below 
$M_{Pl}$, we are  not sensitive to black hole formation and other nonlinear effects 
which could dominate Planck scale physics.

Before closing, we would like to mention that there is a net gravitational 
anomaly for the singlet chiral current associated with the right-handed 
neutrino phases  (gravitational anomalies cancel for 
the Standard Model gauge currents as $\mbox{Tr}(Y)=0 $).   
If the condensate is the only breaking of these 
phases then there should be massless Nambu-Goldstone Bosons (NGBs),  or massive pseudoNGBs if 
there is some other explicit breaking. These NGBs or pseudo NGBs are our would-be
inflatons.  Despite being a
kind of explicit breaking, the gravitational anomaly does not necessarily generate a mass.   
Instantons make the $\eta' $ heavy in QCD but the electroweak anomalies 
make little or no contribution to masses. 
Even if the anomaly  does not generate  a mass for  
the PNGB, it may still have an important role in our model model as it could play a role 
ending the inflationary phase.

\subsection*{ACKNOWLEDGMENTS}
I am deeply grateful  and pleased to acknowledge
Bill Bardeen, with whom I am specially indebted, for his
support, encouragement, and guidance during this work
(and for his invaluable patience with me and my
manuscripts). 
Financial support from Spanish 
MEC and FEDER (EC) under grant FPA2008-02878, and Generalitat 
Valenciana under the grant PROMETEO/2008/004 is also acknowledged. 
Feynman diagrams were made 
using Jaxodraw \cite{Binosi:2003yf}

\appendix

\section{Feynman rules}

We list here the Feyman rules which are employed in our calculations. For  derivations
of these forms, see \cite{Bje01}

The propagator for a massless fermion field  is given by
 \begin{align*}
  \parbox{75pt}{\includegraphics[scale=.75]{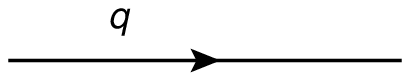}}\qquad &=  \frac{i}{\slashed q}  \; .
 \end{align*}
%
In harmonic (Feynman) gauge the graviton propagator has been discussed in \cite{Don94b,Hoo74}. 
\begin{align*}
\mu \nu  \parbox{79pt}{\includegraphics[scale=.75]{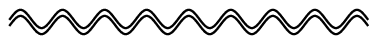}}\alpha \beta 
  \; &  = \frac{i \mathcal P^{\mu\nu\alpha\beta}}{q^2}
\end{align*}
where the polarization sum is given by
\begin{equation*}
 \mathcal P^{\mu\nu\alpha\beta} = \frac{1}{2}\bigl( \eta^{\mu\alpha}\eta^{\nu\beta}+\eta^{\mu\beta}\eta^{\nu\alpha}- \eta^{\mu\nu}\eta^{\alpha\beta}\bigr) \; .
\end{equation*}
%
The 1-graviton-2-(massless)fermion vertex is given by
\begin{align*}
\mu\nu \parbox{60pt}{\includegraphics[scale=.75]{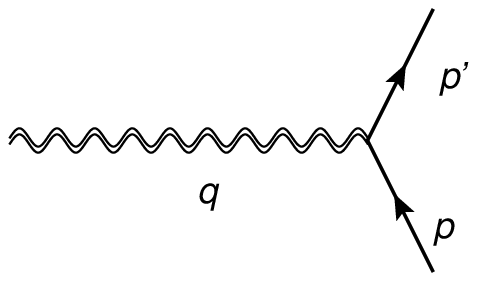}} \qquad \;\;  &= \;\;\tau_1^{\mu\nu} \ ,
\end{align*}
with 
\begin{equation*}
 \; \tau_1^{\mu\nu}= \frac{i \kappa}{2} \left[(p' +p)_\mu \gamma_\nu + (p' +p)_\nu \gamma_\mu 
- \eta_{\mu \nu} ( \slashed p' + \slashed p) \right]
\end{equation*}
%
%
The 1-graviton-2-scalar vertex reads
\begin{align*}
 \mu\nu \parbox{60pt}{\includegraphics[scale=.75]{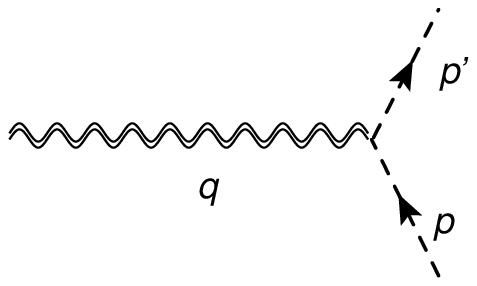}} \qquad \;\; &= \;\;\tau_2^{\mu\nu} \ ,
\end{align*}
with 
\begin{equation*}
 \tau_2^{\mu\nu} = \frac{-i \kappa}{2} \bigl \lbrace p^\mu p^{\prime \, \nu} + p^\nu p^{\prime \, \mu} - \eta^{\mu\nu}  ( p \cdot p' )  \bigr \rbrace
\end{equation*}
%
The 1-graviton-2-photon vertex is given by
\begin{align*}
 \mu \nu \parbox{80pt}{\includegraphics[scale=.75]{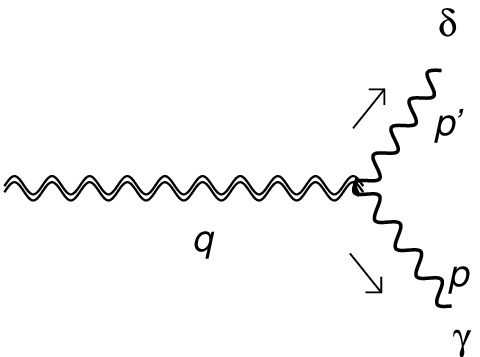}} \qquad \;\; &= 
\;\; \tau_3^{\mu\nu \gamma\delta}  \ ,
\end{align*}
with
\begin{align*}
\tau_3^{\mu\nu \gamma\delta}& = i \kappa \biggl \lbrace \mathcal P^{\mu\nu \gamma \delta} \ (p \cdot p') + \frac{1}{2} \biggl \lbrack \eta^{\mu\nu}p^\delta p^{\prime \, \gamma} 
	+ \eta^{\gamma\delta} \bigl( p^\mu p^{\prime \, \nu} + p^\nu p^{\prime \, \mu} \bigr) - \bigl (\eta^{\mu\delta} p^{\prime \, \gamma} p^\nu + \eta^{\nu\delta} p^{\prime \, \gamma} p^\mu \\
	&+\eta^{\nu\gamma}p^{\prime \,\mu}p^\delta + \eta^{\mu\gamma}p^{\prime \, \nu}p^\delta \bigr)\biggr \rbrack\biggr \rbrace  \ .
\end{align*}
%
Using the background field method the 3-graviton vertex is found as \cite{Don94b}
\begin{align*}
\mu \nu \parbox{80pt}{\includegraphics[scale=.75]{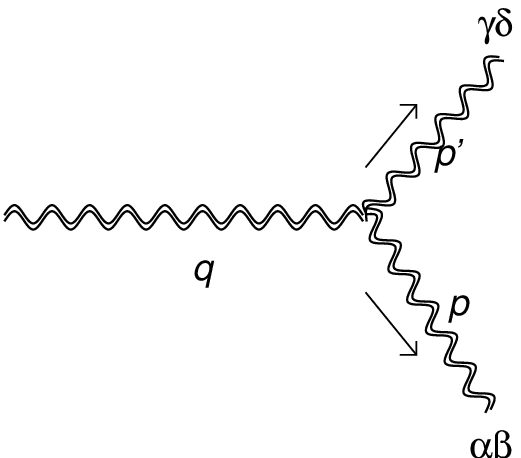}} & \qquad \;\;= 
\;\; \tau^{\mu\nu}_{4 \, \alpha\beta\gamma\delta} (k,q) \ ,
\end{align*}
with
 \begin{equation*}
  \begin{split}
   \tau^{\mu\nu}_{4 \, \alpha\beta\gamma\delta} = -\frac{i \kappa}{2} &\cdot \biggl \lbrace \mathcal P_{\alpha\beta\gamma\delta} \biggl \lbrack k^\mu k^\nu + (k-q)^\mu  (k-q)^\nu + q^\mu q^\nu - \frac{3}{2} \eta^{\mu\nu} q^2 \biggr \rbrack + 2 q_\lambda q_\sigma \bigl \lbrack \tensor{\mathds 1}{_\alpha_\beta^\sigma^\lambda}\ \tensor{\mathds 1}{_\gamma_\delta^\mu^\nu} + \tensor{\mathds 1}{_\gamma_\delta^\sigma^\lambda} \ \tensor{\mathds 1}{_\alpha_\beta^\mu^\nu} - \tensor{\mathds 1}{_\alpha_\beta^\mu^\sigma} \ \tensor{\mathds 1}{_\gamma_\delta^\nu^\lambda} \\
	& - \tensor{\mathds 1}{_\gamma_\delta^\mu^\sigma} \ \tensor{\mathds 1}{_\alpha_\beta^\nu^\lambda} \bigr \rbrack+ \bigl \lbrack q_\lambda q^\mu \bigr( \eta_{\alpha\beta} \tensor{\mathds 1}{_\gamma_\delta^\nu^\lambda} + \eta_{\gamma\delta} \tensor{\mathds 1}{_\alpha_\beta^\nu^\lambda} \bigr) +q_\lambda q^\nu \bigr( \eta_{\alpha\beta} \tensor{\mathds 1}{_\gamma_\delta^\mu^\lambda} + \eta_{\gamma\delta} \tensor{\mathds 1}{_\alpha_\beta^\mu^\lambda} \bigr)\\
	&-q^2 \bigl( \eta_{\alpha\beta} \tensor{\mathds 1}{_\gamma_\delta^\mu^\nu}+ \eta_{\gamma\delta}\tensor{\mathds 1}{_\alpha_\beta^\mu^\nu}\bigr)- \eta^{\mu\nu} q_\lambda q_\sigma \bigr( \eta_{\alpha\beta}\tensor{\mathds 1}{_\gamma_\delta^\sigma^\lambda} + \eta_{\gamma\delta} \tensor{\mathds 1}{_\alpha_\beta^\sigma^\lambda} \bigr) \bigr \rbrack \\
	&+ \bigl \lbrack 2 q_\lambda \bigl \lbrace \tensor{\mathds 1}{_\alpha_\beta^\lambda^\sigma} \ \tensor{\mathds 1}{_\gamma_\delta_\sigma^\nu} (k-q)^\mu + \tensor{\mathds 1}{_\alpha_\beta^\lambda^\sigma} \ \tensor{\mathds 1}{_\gamma_\delta_\sigma^\mu} (k-q)^\nu - \tensor{\mathds 1}{_\gamma_\delta^\lambda^\sigma} \ \tensor{\mathds 1}{_\alpha_\beta_\sigma^\nu} k^\mu - \tensor{\mathds 1}{_\gamma_\delta^\lambda^\sigma} \ \tensor{\mathds 1}{_\alpha_\beta_\sigma^\mu} k^\nu \bigr \rbrace \\
	& + q^2 \bigl( \tensor{\mathds 1}{_\alpha_\beta_\sigma^\mu}\tensor{\mathds 1}{_\gamma_\delta^\nu^\sigma} + \tensor{\mathds 1}{_\alpha_\beta^\nu^\sigma}\tensor{\mathds 1}{_\gamma_\delta_\sigma^\mu} \bigr) + \eta^{\mu\nu} q_\sigma q_\lambda \bigl( \tensor{\mathds 1}{_\alpha_\beta^\lambda^\rho} \ \tensor{\mathds 1}{_\gamma_\delta_\rho^\sigma}  + \tensor{\mathds 1}{_\gamma_\delta^\lambda^\rho} \ \tensor{\mathds 1}{_\alpha_\beta_\rho^\sigma} \bigr) \bigr \rbrack + \pmb{\biggl \lbrace} \bigl( k^2 + (k-q)^2 \bigr) \\
	 &\cdot \biggl \lbrack \tensor{\mathds 1}{_\alpha_\beta^\mu^\sigma} \ \tensor{\mathds 1}{_\gamma_\delta_\sigma^\nu} +\tensor{\mathds 1}{_\gamma_\delta^\mu^\sigma} \ \tensor{\mathds 1}{_\alpha_\beta_\sigma^\nu} - \frac{1}{2} \eta^{\mu\nu} \mathcal P_{\alpha\beta\gamma\delta} \biggr \rbrack - \bigl( \tensor{\mathds 1}{_\gamma_\delta^\mu^\nu} \ \eta_{\alpha\beta} \ k^2 - \tensor{\mathds 1}{_\alpha_\beta^\mu^\nu} \ \eta_{\gamma\delta} \ (k-q)^2 \bigr) \pmb{\biggr \rbrace} \biggr \rbrace \; .
  \end{split}
 \end{equation*}
and
\begin{equation*}
 \mathds 1_{\alpha\beta\gamma\delta} = \frac{1}{2}\bigl( \eta_{\alpha\gamma}\eta_{\beta\delta}+ \eta_{\alpha\delta}\eta_{\beta\gamma}\bigr) \ .
\end{equation*}
%

a complete set of Feynman rules can be found in \cite{Bje03,Faller:2007sy}.

\end{document}